\documentclass[aps,prd,twocolumn,superscriptaddress]{revtex4}
\usepackage{epsfig,epsf}
\usepackage{amsmath}
\usepackage{amsthm}
\usepackage{amsfonts}
\usepackage{amssymb}
\usepackage{dsfont}
\usepackage{multirow}
\usepackage{appendix}
\usepackage{slashed}
\usepackage[active]{srcltx}
\usepackage{psfrag}

\newcommand{\func}[1]{\operatorname{#1}}

\begin{document}

\title{Double-heavy axial-vector tetraquark $T_{bc;\bar{u}\bar{d}}^{0}$ }
\author{S.~S.~Agaev}
\affiliation{Institute for Physical Problems, Baku State University, Az--1148 Baku,
Azerbaijan}
\author{K.~Azizi}
\thanks{Corresponding author}
\affiliation{Department of Physics, University of Tehran, North Karegar Ave., Tehran
14395-547, Iran}
\affiliation{Department of Physics, Do\v{g}u\c{s} University, Acibadem-Kadik\"{o}y, 34722
Istanbul, Turkey}
\author{H.~Sundu}
\affiliation{Department of Physics, Kocaeli University, 41380 Izmit, Turkey}

\begin{abstract}
The mass and coupling of the axial-vector tetraquark $T_{bc;\bar{u}\bar{d}%
}^{0}$ (in a short form $T_{bc}^{0}$) are calculated by means of the QCD
two-point sum rule method. In computations we take into account
contributions arising from various quark, gluon and mixed vacuum condensates
up to dimension 10. The central value of the mass $m=(7105 \pm 155)~\mathrm{%
MeV}$ lies below the thresholds for the strong and electromagnetic decays of
$T_{bc}^{0}$ state, and hence it transforms to conventional mesons only
through the weak decays. In the case of $m=7260~\mathrm{MeV}$ the tetraquark
$T_{bc}^{0}$ becomes the strong- and electromagnetic-interaction unstable
particle. In the first case, we find the full width and mean lifetime of $%
T_{bc}^{0}$ using its dominant semileptonic decays $T_{bc}^{0}\to T_{cc;\bar{%
u}\bar{d}}^{+}l\overline{\nu }_{l}$ ($l=e,\ \mu, \tau$), where the
final-state tetraquark is a scalar state. We compute also partial widths of
the nonleptonic weak decays $T_{bc}^{0}\to T_{cc;\bar{u}\bar{d}%
}^{+}\pi^{-}(K^{-}, D^{-}, D_{s}^{-})$, and take into account their effects
on the full width of $T_{bc}^{0}$. In the context of the second scenario we
calculate partial widths of $S$-wave strong decays $T_{bc}^{0}\to B^{\ast
-}D^{+}$ and $T_{bc}^{0}\to \overline{B}^{\ast 0}D^{0}$, and using these
channels evaluate the full width of $T_{bc}^{0}$. Predictions for $\Gamma _{%
\mathrm{full}} =(3.98\pm 0.51)\times 10^{-10}~\mathrm{MeV}$ and mean
lifetime $\tau=1.65_{-0.18}^{+0.25}~\mathrm{ps}$ of $T_{bc}^{0}$ obtained in
the context of the first option, as well as the full width $\Gamma_{\mathrm{%
full}}=(63.5\pm 8.9)~\mathrm{MeV}$ extracted in the second scenario may be
useful for experimental and theoretical exploration of double-heavy exotic
mesons.
\end{abstract}

\maketitle


\section{Introduction}

\label{sec:Int}
During last two decades double-heavy tetraquarks as real candidates for
stable four-quark states became subjects of intensive studies. In the
pioneering papers \cite{Ader:1981db,Lipkin:1986dw,Zouzou:1986qh} it was
demonstrated that a heavy $Q$ and light $q$ quarks may form the stable
exotic mesons $QQ\bar{q}\bar{q}$ provided the ratio $m_{Q}/m_{q}$ is large
enough. These results were obtained in the context of a potential model with
the additive pairwise interaction, but even models with relaxed restrictions
on the confining potential led to the similar predictions. Indeed, in
accordance with Ref. \cite{Carlson:1987hh} the isoscalar axial-vector
tetraquark $T_{bb;\overline{u}\overline{d}}^{-}$ turns to be
strong-interaction stable state that lies below the $B\overline{B}^{\ast }$
threshold. It is worth noting that an only constraint imposed in Ref. \cite%
{Carlson:1987hh} on the potential was its finiteness at close distances of
two particles. Therefore, $T_{bb;\overline{u}\overline{d}}^{-}$ decays to
conventional mesons only through weak processes and has a long lifetime,
which is important for its experimental exploration. A situation with the
tetraquarks $T_{bc;\bar{q}\bar{q}^{\prime }}$ and $T_{cc;\bar{q}\bar{q}%
^{\prime }}$ was not clear, because $bc$ and $cc$ diquarks might constitute
both stable and unstable states.

In years followed after this progress, various models of high energy physics
were used to investigate the double-heavy tetraquarks $T_{QQ}$ \cite%
{Janc:2004qn,Cui:2006mp,Vijande:2006jf,Ebert:2007rn,Navarra:2007yw,Du:2012wp,Hyodo:2012pm,Esposito:2013fma}%
. Recent interest to these problems was inspired by results of the LHCb
Collaboration on properties of the doubly charmed baryon $\Xi _{cc}^{++}=ccu$
\cite{Aaij:2017ueg}. Parameters of this baryon were used in Ref. \cite%
{Karliner:2017qjm} to evaluate the mass and analyze possible decay channels
of $T_{bb;\overline{u}\overline{d}}^{-}$. Predictions obtained there
confirmed the stability of $T_{bb;\overline{u}\overline{d}}^{-}$ against the
strong and electromagnetic decays to $B^{-}\overline{B}^{\ast 0}$ and $B^{-}%
\overline{B}^{0}\gamma $, respectively. The strong-interaction stable nature
of the tetraquarks $T_{bb;\overline{u}\overline{d}}^{-}$, $T_{bb;\overline{u}%
\overline{s}}^{-}$, and $T_{bb;\overline{d}\overline{s}}^{0}$ was
demonstrated in Ref.\ \cite{Eichten:2017ffp} by invoking heavy-quark
symmetry relations. The mass and coupling of $T_{bb;\overline{u}\overline{d}%
}^{-}$ was evaluated in our work \cite{Agaev:2018khe} as well, in which we
estimated also its full width and mean lifetime using the semileptonic decay
channel $T_{bb;\overline{u}\overline{d}}^{-}\rightarrow Z_{bc;\overline{u}%
\overline{d}}^{0}l\bar{\nu _{l}}$ .

Another class of four-quark mesons, namely one that contains the heavy
diquarks $bc$ is on agenda of physicists as well. The scalar and
axial-vector tetraquarks $bc\overline{u}\overline{d}$ are particles of
special interest, because they may form strong-interaction stable compounds.
But calculations performed in the context of different approaches lead
controversial results. Thus, the Bethe-Salpeter method predicts the mass of
the scalar tetraquark $Z_{bc;\overline{u}\overline{d}}^{0}$ (in what follows
$Z_{bc}^{0}$) at around $6.93~\mathrm{GeV}$, which is below the threshold $%
7145~\mathrm{MeV}$ for $S$-wave strong decays to heavy mesons $B^{-}D^{+}$
and $\overline{B^{0}}D^{0}$ \cite{Feng:2013kea}. Recent analysis
demonstrated that $Z_{bc}^{0}$ lies $11~\mathrm{MeV}$ below this threshold
\cite{Karliner:2017qjm}, whereas the authors of Ref.\ \cite{Eichten:2017ffp}
found the masses of the scalar and axial-vector tetraquarks $bc\overline{u}%
\overline{d}$ equal to $7229~\mathrm{MeV}$ and $7272~\mathrm{MeV}$,
respectively. These predictions make kinematically allowed their strong
decays to ordinary $B^{-}D^{+}/\overline{B}^{0}D^{0}$ and $B^{\ast }D$
mesons.

It is interesting that lattice calculations prove the strong-interaction
stabile nature of the axial-vector tetraquark $ud\overline{b}\overline{c}$,
because its mass is below the $\overline{D}B^{\ast }$ threshold \cite%
{Francis:2018jyb}. However, the authors could not decide would this exotic
meson decay weakly or might transform also to the final state $\overline{D}%
B\gamma $. The stability of $J^{P}=0^{+}$ and $1^{+}$ isoscalar tetraquarks $%
bc\overline{u}\overline{d}$ was confirmed in Ref.\ \cite{Caramees:2018oue},
in which it was found that $J^{P}=0^{+}$ state is a strong- and
electromagnetic-interaction stable particle, whereas $J^{P}=1^{+}$ may also
transform through the electromagnetic interaction.

In the context of the QCD sum rule approach the spectroscopic parameters of
the scalar tetraquark $Z_{bc}^{0}$ were calculated also in our work \cite%
{Agaev:2018khe}. For the mass of $Z_{bc}^{0}$ computations predicted $%
m_{Z}=(6660\pm 150)~\mathrm{MeV}$, which is considerably below the threshold
$7145~\mathrm{MeV}$. The electromagnetic decay modes $Z_{bc}^{0}\rightarrow
\overline{B}^{0}D_{1}^{0}\gamma $ and $B^{\ast }D_{0}^{\ast }\gamma $ are
among forbidden processes as well, because relevant thresholds exceed $7600~%
\mathrm{MeV}$ and are higher than the mass of $Z_{bc}^{0}$. In other words,
in accordance with our results the scalar tetraquark $Z_{bc}^{0}$ is a
strong- and electromagnetic-interaction stable particle. The $Z_{bc}^{0}$
transforms due to weak decays, which allowed us to find in Ref.\ \cite%
{Sundu:2019feu} its full width and mean lifetime.

In the present article we study the axial-vector tetraquark $T_{bc;\overline{%
u}\overline{d}}^{0}$ (hereafter $T_{bc}^{0}$) by computing its spectroscopic
parameters, full width and mean lifetime. The mass $m$ and coupling $f$ of $%
T_{bc}^{0}$ are evaluated in the framework of the QCD two-point sum rule
method by taking into account vacuum expectation values of the local quark,
gluon and mixed operators up to dimension ten. The mass of $T_{bc}^{0}$
extracted in the present work $m=(7105\pm 155)~\mathrm{MeV}$ contains
theoretical errors typical for sum rule computations, hence, there are two
options to find its full width and estimate mean lifetime. Thus, the central
value of the mass is lower than the thresholds $7190~\mathrm{MeV}$ and $7286~%
\mathrm{MeV}$ for strong $S$-wave decays of $T_{bc}^{0}$ to final states $%
B^{\ast -}D^{+}/\overline{B}^{\ast 0}D^{0}$ and $B^{-}D^{\ast +}/\overline{B}%
^{0}D^{\ast 0}$, respectively. This mass is also lower than the threshold $%
7145~\mathrm{MeV}$ for the electromagnetic decays $D^{+}B^{-}\gamma /D^{0}%
\overline{B}^{0}\gamma $. Therefore, in this case the full width and
lifetime of the exotic meson $T_{bc}^{0}$ should be determined from its weak
decays. But considering the maximum theoretical prediction for $m=7260$ $%
\mathrm{MeV}$, one sees that it is higher than the threshold for strong
decays $B^{\ast -}D^{+}/\overline{B}^{\ast 0}D^{0}$ and electromagnetic
transitions $D^{+}B^{-}\gamma /D^{0}\overline{B}^{0}\gamma $. Realization of
this scenario means that the width of the tetraquark $T_{bc}^{0}$ is
determined mainly by strong decays, because partial widths of weak and
electromagnetic processes are very small and can be neglected.

Here, to calculate the full width of the tetraquark $T_{bc}^{0}$, we
consider both scenarios. In the first case $m=7105~\mathrm{MeV}$, and the
processes $T_{bc}^{0}\rightarrow T_{cc;\overline{u}\overline{d}}^{+}l%
\overline{\nu }_{l}$ ($l=e,\mu $ and $\tau $), where the final-state
tetraquark $T_{cc;\overline{u}\overline{d}}^{+}$ (in what follows $%
T_{cc}^{+} $ ) is a scalar particle, are the dominant semileptonic decay
channels of $T_{bc}^{0}$. These decays run due to transition $b\rightarrow
W^{-}c$. The differential rates of these semileptonic decays are determined
by the weak form factors $G_{i}(q^{2})$ ($i=0,1,2,3$), which are evaluated
by employing the QCD three-point sum rule approach. Then, partial width of
the processes $T_{bc}^{0}\rightarrow T_{cc}^{+}l\overline{\nu }_{l}$ can be
found by integrating the relevant differential rates over the momentum
transfer $q^{2} $. The sum rule method does not encompass all kinematically
allowed values of $q^{2}$, therefore we introduce fit functions that
coincide with sum rule predictions, and can be extrapolated to cover a whole
integration region.

But a decay $b\rightarrow W^{-}c$ can be followed by transitions $%
W^{-}\rightarrow d\overline{u},\ s\overline{u},\ d\overline{c}$ and $s%
\overline{c}$ as well. Afterwards these quark pairs can form ordinary mesons
through different mechanisms. Thus, in the hard-scattering picture a pair $d%
\overline{u}$ , for example, can create conventional mesons with $q\overline{%
q}$ quarks appeared due to a gluon from one of $d$ or $\overline{u}$ quarks.
These processes generate final states $T_{bc}^{0}\rightarrow
T_{cc}^{+}M_{1}(d\overline{q})M_{2}(q\overline{u})$ which are suppressed
relative to the semileptonic decays by the factor $\alpha
_{s}^{2}|V_{ud}|^{2}$. Alternatively, pairs of quarks $d\overline{u},\ s%
\overline{u},\ d\overline{c}$ and $s\overline{c}$ can form $\pi ^{-}$, $K^{-%
\text{ }}$, $D^{-}$ and $D_{s}^{-}$ mesons triggering the two-body
nonleptonic decays $T_{bc}^{0}\rightarrow T_{cc}^{+}\pi ^{-}(K^{-\text{ }%
},D^{-},D_{s}^{-})$. Another class of the $T_{bc}^{0}$ tetraquark's weak
decays is connected with possibility of direct combination of these quarks
with ones from $T_{cc;\overline{u}\overline{d}}^{+}$ and creation of
three-meson final states. The two-body and three-meson nonleptonic decays do
not suppressed by additional factors relative to the semileptonic decays,
and their contributions to full width of $T_{bc}^{0}$ may be considerable.

In the second scenario $m=7260~\mathrm{MeV}$, and this mass is above the
threshold for strong decays to mesons $B^{\ast -}D^{+}/\overline{B}^{\ast
0}D^{0}$, but is still below the threshold for other two possible decay
modes to final states $B^{-}D^{\ast +}/\overline{B}^{0}D^{\ast 0}$.
Therefore, we calculate the partial width of the kinematically allowed
strong $S$-wave decays $T_{bc}^{0}\rightarrow B^{\ast -}D^{+}$ and $%
T_{bc}^{0}\rightarrow \overline{B}^{\ast 0}D^{0}$. To this end, we use again
the QCD three-point sum rule method and evaluate the strong form factors $%
g_{1}(q^{2})$ and $g_{2}(q^{2})$. By extrapolating these form factors to the
corresponding mass shells we determine couplings of the vertices $%
T_{bc}^{0}B^{\ast -}D^{+}$ and $T_{bc}^{0}\overline{B}^{\ast 0}D^{0}$, and
calculate partial width of these decays. The full width of the tetraquark $%
T_{bc}^{0}$ is evaluated using these two dominant strong decay channels.

This article is organized in the following manner: In Section \ref{sec:Mass}%
, from analysis of the two-point correlation function with an appropriate
interpolating current, we derive sum rules to evaluate the spectroscopic
parameters of the tetraquark $T_{bc}^{0}$. In the next Section \ref%
{sec:Decays1}, using the parameters of $T_{bc}^{0}$ and ones of the
final-state tetraquark, we calculate the partial width of its dominant
semileptonic decays. To this end, we derive the sum rules for the weak form
factors and by means of fit functions extrapolate them to the whole region,
where an integration over $q^{2}$ should be carried out. In Section \ref%
{sec:Decays2}, we analyze the nonleptonic weak decays $T_{bc}^{0}\rightarrow
T_{cc}^{+}\pi ^{-}(K^{-\text{ }},D^{-},D_{s}^{-})$ of the tetraquark $%
T_{bc}^{0}$ and find their partial widths. Here, we also calculate the full
width of $T_{bc}^{0}$ in the first scenario, i.e., for $m=7105~\mathrm{MeV}$%
. The Sec. \ref{sec:Decays3} is devoted to calculation of the partial widths
of the strong processes $T_{bc}^{0}\rightarrow B^{\ast -}D^{+}$ and $%
T_{bc}^{0}\rightarrow \overline{B}^{\ast 0}D^{0}$, \ where we also evaluate
the full width of the tetraquark $T_{bc}^{0}$ if $m=7260~\mathrm{MeV}$.
Section \ref{sec:Disc} is reserved for analysis of obtained results, and
contains also our concluding notes.


\section{Mass and coupling of the axial-vector tetraquark $T_{bc}^{0}$}

\label{sec:Mass}
In this section we extract the spectroscopic parameters of the axial-vector
tetraquark $T_{bc}^{0}$ from the QCD sum rules. To this end, we start from
analysis of the correlation function $\Pi _{\mu \nu }(p)$, which is given by
the formula

\begin{equation}
\Pi _{\mu \nu }(p)=i\int d^{4}xe^{ipx}\langle 0|\mathcal{T}\{J_{\mu
}(x)J_{\nu }^{\dag }(0)\}|0\rangle .  \label{eq:CF1}
\end{equation}%
Here $J_{\mu }(x)$ is the interpolating current to the axial-vector
tetraquark $T_{bc}^{0}$. We suggest that $T_{bc}^{0}$ is built of the scalar
diquark and axial-vector antidiquark, and hence its current has the form
\begin{eqnarray}
J_{\mu }(x) &=&b_{a}^{T}(x)C\gamma _{5}c_{b}(x)\left[ \overline{u}%
_{a}(x)\gamma _{\mu }C\overline{d}_{b}^{T}(x)\right.  \notag \\
&&\left. -\overline{u}_{b}(x)\gamma _{\mu }C\overline{d}_{a}^{T}(x)\right] .
\label{eq:Curr}
\end{eqnarray}%
Here $a$ and $b$ are the color indices and $C$ is the charge conjugation
operator. The current (\ref{eq:Curr}) has the antisymmetric color structure $%
[\overline{3}_{c}]_{bc}\otimes \lbrack 3_{c}]_{\overline{u}\overline{d}}$
and describes a four-quark state with the quantum numbers $1^{+}$, where $%
b^{T}C\gamma _{5}c$ and $\overline{u}\gamma _{\mu }C\overline{d}^{T}$ are
the scalar diquark and axial-vector antidiquark, respectively.

To derive required sum rules we find, in accordance with prescriptions of
the method, the correlation function $\Pi _{\mu \nu }(p)$ using the
tetraquark's mass $m$ and coupling $f$. We consider it as a ground-state
particle, and isolate the first term in $\Pi _{\mu \nu }^{\mathrm{Phys}}(p)$
\begin{equation}
\Pi _{\mu \nu }^{\mathrm{Phys}}(p)=\frac{\langle 0|J_{\mu
}|T_{bc}^{0}(p)\rangle \langle T_{bc}^{0}(p)|J_{\nu }^{\dagger }|0\rangle }{%
m^{2}-p^{2}}+\dots .  \label{eq:CF2}
\end{equation}%
Equation (\ref{eq:CF2}) is obtained by saturating the correlation function
with a complete set of $J^{P}=1^{+}$ states and carrying out the integration
over $x$. Contributions of higher resonances and continuum states to $\Pi
_{\mu \nu }^{\mathrm{Phys}}(p)$ are denoted by the dots.

To simplify further the correlator $\Pi _{\mu \nu }^{\mathrm{Phys}}(p)$ it
is useful to define the matrix element
\begin{equation}
\langle 0|J_{\mu }|T_{bc}^{0}(p,\epsilon )\rangle =fm\epsilon _{\mu },
\label{eq:MElem1}
\end{equation}%
with $\epsilon _{\mu }$ being the polarization vector of the $T_{bc}^{0}$
state. Then in terms of $m$ and $f$ \ the correlation function $\Pi _{\mu
\nu }^{\mathrm{Phys}}(p)$ takes the form
\begin{equation}
\Pi _{\mu \nu }^{\mathrm{Phys}}(p)=\frac{m^{2}f^{2}}{m^{2}-p^{2}}\left(
-g_{\mu \nu }+\frac{p_{\mu }p_{\nu }}{m^{2}}\right) +\ldots .
\label{eq:CorM}
\end{equation}

The QCD side of the sum rule is determined by the correlation function $\Pi
_{\mu \nu }(p)$, but calculated now by employing the quark propagators
\begin{eqnarray}
&&\Pi _{\mu \nu }^{\mathrm{OPE}}(p)=i\int d^{4}xe^{ipx}\left. \mathrm{Tr}%
\left[ \gamma _{5}\widetilde{S}_{b}^{aa^{\prime }}(x)\gamma
_{5}S_{c}^{bb^{\prime }}(x)\right] \right.  \notag \\
&&\times \left\{ \mathrm{Tr}\left[ \gamma _{\mu }\widetilde{S}%
_{d}^{a^{\prime }b}(-x)\gamma _{\nu }S_{u}^{b^{\prime }a}(-x)\right] -%
\mathrm{Tr}\left[ \gamma _{\mu }\widetilde{S}_{d}^{b^{\prime }b}(-x)\right.
\right.  \notag \\
&&\left. \times \gamma _{\nu }S_{u}^{a^{\prime }a}(-x)\right] -\mathrm{Tr}%
\left[ \gamma _{\mu }\widetilde{S}_{d}^{a^{\prime }a}(-x)\gamma _{\nu
}S_{u}^{b^{\prime }b}(-x)\right]  \notag \\
&&\left. +\mathrm{Tr}\left[ \gamma _{\mu }\widetilde{S}_{d}^{b^{\prime
}a}(-x)\gamma _{\nu }S_{u}^{a^{\prime }b}(-x)\right] \right\} ,
\label{eq:CF3}
\end{eqnarray}%
where $S_{q}^{ab}(x)$ is the heavy $(b,c)$- or light $(u,d)$-quark
propagators. Their explicit expressions can be found in Ref.\ \cite%
{Sundu:2018uyi}. In Eq.\ (\ref{eq:CF3}) we use the shorthand notation
\begin{equation}
\widetilde{S}_{q}(x)=CS_{q}^{T}(x)C.  \label{eq:Prop}
\end{equation}

The correlation function $\Pi _{\mu \nu }(p)$ contains the different Lorentz
structures one of which should be chosen to get the sum rules. The invariant
amplitudes $\Pi ^{\mathrm{Phys}}(p^{2})$ and $\Pi ^{\mathrm{OPE}}(p^{2})$
corresponding to the terms $\sim g_{\mu \nu }$ are convenient for our aim,
because they do not receive contributions from the scalar particles.

After picking up and equating corresponding invariant amplitudes, we apply
the Borel transformation to both sides of the obtained expression. This is
necessary to suppress contributions of the higher resonances and continuum
states. Afterwards, one has to subtract continuum contributions, which is
achieved by invoking suggestion on the quark-hadron duality. The obtained
equality acquires a dependence on auxiliary parameters of the sum rules $%
M^{2}$ and $s_{0}$: first of them is the Borel parameter appeared due to
corresponding transformation, the second one $s_{0}$ is the continuum
subtraction parameter that separates the ground-state and higher resonances
from each another.

The final sum rule for the mass of the state $T_{bc}^{0}$ reads:
\begin{equation}
m^{2}=\frac{\int_{\mathcal{M}^{2}}^{s_{0}}dss\rho ^{\mathrm{OPE}%
}(s)e^{-s/M^{2}}}{\int_{\mathcal{M}^{2}}^{s_{0}}ds\rho ^{\mathrm{OPE}%
}(s)e^{-s/M^{2}}},  \label{eq:Mass1}
\end{equation}%
where $\mathcal{M}=m_{b}+m_{c}$. For the coupling $f$ one obtains the
expression
\begin{equation}
f^{2}=\frac{1}{m^{2}}\int_{\mathcal{M}^{2}}^{s_{0}}ds\rho ^{\mathrm{OPE}%
}(s)e^{(m^{2}-s)/M^{2}}.  \label{eq:Coupl1}
\end{equation}%
Here $\rho ^{\mathrm{OPE}}(s)$ is the two-point spectral density, which is
determined as an imaginary part of the term in $\Pi _{\mu \nu }^{\mathrm{OPE}%
}(p)$ proportional to $g_{\mu \nu }$, and calculated by taking into account
the quark, gluon and mixed vacuum condensates up to dimension ten. Explicit
expression of $\rho ^{\mathrm{OPE}}(s)$ is rather cumbersome, hence we
refrain from providing it here.

In addition to $M^{2}$ and $s_{0}$, numerical values of which depend on the
considering problem, the sum rules (\ref{eq:Mass1}) and (\ref{eq:Coupl1})
contain also the vacuum condensates, as well as the masses of $b$ and $c$%
-quarks
\begin{eqnarray}
&&\langle \bar{q}q\rangle =-(0.24\pm 0.01)^{3}~\mathrm{GeV}^{3},\   \notag \\
&&\langle \overline{q}g_{s}\sigma Gq\rangle =m_{0}^{2}\langle \overline{q}%
q\rangle ,\ m_{0}^{2}=(0.8\pm 0.1)~\mathrm{GeV}^{2},  \notag \\
&&\langle \frac{\alpha _{s}G^{2}}{\pi }\rangle =(0.012\pm 0.004)~\mathrm{GeV}%
^{4},  \notag \\
&&\langle g_{s}^{3}G^{3}\rangle =(0.57\pm 0.29)~\mathrm{GeV}^{6},  \notag \\
&&m_{b}=4.18_{-0.03}^{+0.04}~\mathrm{GeV},\ m_{c}=1.275_{-0.035}^{+0.025}\
\mathrm{GeV}.  \label{eq:Parameters}
\end{eqnarray}

The parameters $M^{2}$ and $s_{0}$ should satisfy constraints that are
standard for the the sum rule computations. Thus, at maximum of the Borel
parameter the pole contribution ($\mathrm{PC}$) should be larger than some
fixed value, whereas the main criterium to fix the minimum of a Borel window
is convergence of the operator product expansion (OPE). Additionally, at
minimum $M^{2}$ the perturbative contribution has to exceed the
nonperturbative terms considerably. Because quantities extracted from the
sum rules demonstrate dependence on the auxiliary parameters, the regions
for $M^{2}$ and $s_{0}$ should minimize these side effects, as well.
\begin{figure}[h]
\includegraphics[width=8.8cm]{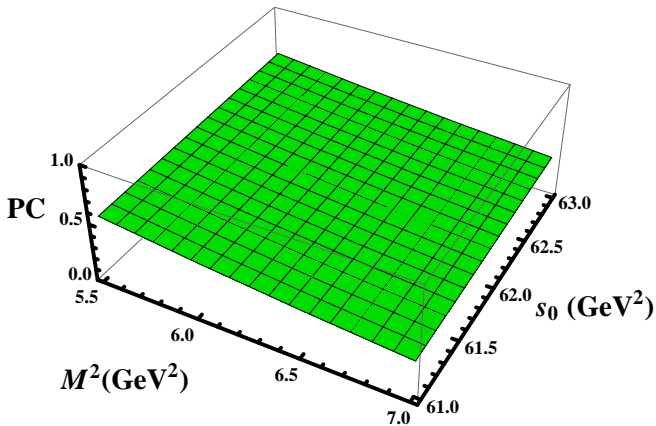}
\caption{ The pole contribution as a function of the Borel and continuum
threshold parameters $M^{2}$ and $s_{0}$.}
\label{fig:PC}
\end{figure}

\begin{figure}[h]
\includegraphics[width=8.8cm]{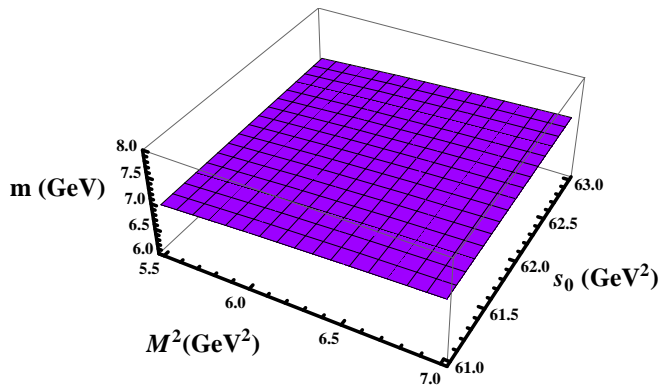}
\caption{ The same as in Fig.\ 1, but for the mass of the tetraquark $%
T_{bc}^{0}$. }
\label{fig:Mass}
\end{figure}

Our analysis proves that the working regions
\begin{equation}
M^{2}\in \lbrack 5.5,~7]~\mathrm{GeV}^{2},\ s_{0}\in \lbrack 61,~63]~\mathrm{%
GeV}^{2},  \label{eq:Reg1}
\end{equation}%
satisfy all aforementioned restrictions. Thus, within the region $M^{2}\in
\lbrack 5.5,~7]~\mathrm{GeV}^{2}$ the pole contribution decreases
approximately from $58\%$ till $34\%$. A detailed picture for $\mathrm{PC}$
is presented in Fig.\ \ref{fig:PC}, where we plot the pole contribution as a
function of $M^{2}$ and $s_{0}$. The minimum $M_{\min }^{2}$ is found from
analysis of the ratio
\begin{equation}
R(M^{2})=\frac{\Pi ^{\mathrm{DimN}}(M^{2},\ s_{0})}{\Pi (M^{2},\ s_{0})},
\label{eq:Conv}
\end{equation}%
where $\Pi (M^{2},s_{0})$ is the Borel transformed and subtracted function $%
\Pi ^{\mathrm{OPE}}(p^{2})$. In the present work as a measure of the
convergence we use the sum of last three terms in OPE $\mathrm{DimN}=\mathrm{%
Dim(8+9+10)}$ and impose the constraint on $R(M^{2})$: the restriction $%
R(M_{\min }^{2})$ $\leq 0.01$ is fulfilled at $5.5\ \mathrm{GeV}^{2}$. The
perturbative contribution at $M^{2}=5.5~\mathrm{GeV}^{2}$ amounts to $68\%$
of the full result and overshoots contribution of the nonperturbative terms.
In Fig.\ \ref{fig:Mass} we demonstrate the dependence of the mass $m$ on $%
M^{2}$ and $s_{0}$, where weak residual effects of these parameters are seen.

Our results for $m$ and $f$ read:
\begin{eqnarray}
m &=&(7105~\pm 155)~\mathrm{MeV},  \notag \\
f &=&(1.0\pm 0.2)\times 10^{-2}~\mathrm{GeV}^{4}.  \label{eq:CMass}
\end{eqnarray}%
Theoretical errors of the mass is milder than ones of the coupling,
nevertheless all these ambiguities do not exceed standard limits of sum rule
computations reaching $2.2\%$ and $\pm 20\%$ of the corresponding central
values, respectively.The spectroscopic parameters of the axial-vector
tetraquark $T_{bc}^{0}$ evaluated in this section form a basis for our
further investigations.

\section{Semileptonic decays $T_{bc}^{0}\rightarrow T_{cc}^{+}l\overline{%
\protect\nu }_{l}$}

\label{sec:Decays1}
As it has been emphasized above for $m=7105~\mathrm{MeV}$ the tetraquark $%
T_{bc}^{0}$ is stable against the strong and electromagnetic interactions,
because then $m$ resides $85/190~\mathrm{MeV}$ and $45~\mathrm{MeV}$ below
the strong and electromagnetic thresholds, respectively. The semileptonic
decays $T_{bc}^{0}\rightarrow T_{cc}^{+}l\overline{\nu }_{l}$ of the
tetraquark $T_{bc}^{0}$ are caused by weak transition $b\rightarrow
W^{-}c\rightarrow cl\overline{\nu }$ of the heavy $b$-quark. It is not
difficult to see, that due to large mass difference between the tetraquarks $%
T_{bc}^{0}$ and $T_{cc}^{+}$, all of the transitions $T_{bc}^{0}\rightarrow
T_{cc}^{+}l\overline{\nu }_{l}$ with $l=e,\ \mu $ and $\tau $ are
kinematically allowed processes. We restrict ourselves by considering only
the dominant process $b\rightarrow W^{-}c$, because due to smallness of the
Cabibbo-Kobayashi-Maskawa (CKM) matrix element $|V_{bu}|^{2}/|V_{bc}|^{2}$ $%
\simeq 0.01$ the decay $b\rightarrow W^{-}u$ is suppressed relative to the
first one.

At the tree-level, the transition $b\rightarrow W^{-}c$ is described by
means of the effective Hamiltonian
\begin{equation}
\mathcal{H}^{\mathrm{eff}}=\frac{G_{F}}{\sqrt{2}}V_{bc}\overline{c}\gamma
_{\mu }(1-\gamma _{5})b\overline{l}\gamma ^{\mu }(1-\gamma _{5})\nu _{l}.
\label{eq:EffecH}
\end{equation}%
Here $G_{F}$ is the Fermi coupling constant, and $V_{bc}$ is the element of
the CKM matrix. After substituting $\mathcal{H}^{\mathrm{eff}}$ between the
initial and final tetraquark fields and factoring out the leptonic piece we
get the matrix element of the current%
\begin{equation}
J_{\mu }^{\mathrm{tr}}=\overline{c}\gamma _{\mu }(1-\gamma _{5})b,
\label{eq:TrCurr}
\end{equation}%
which has to be calculated in terms of the weak form factors $G_{i}(q^{2})$:
they parameterize the long-distance dynamics of the transition
\begin{eqnarray}
&&\langle T_{cc}^{+}(p^{\prime })|J_{\mu }^{\mathrm{tr}}|T_{bc}^{0}(p,%
\epsilon )\rangle =\overline{m}G_{0}(q^{2})\epsilon _{\mu }+\frac{%
G_{1}(q^{2})}{\overline{m}}(\epsilon p^{\prime })P_{\mu }  \notag \\
&&+\frac{G_{2}(q^{2})}{\overline{m}}(\epsilon p^{\prime })q_{\mu }+i\frac{%
G_{3}(q^{2})}{\overline{m}}\varepsilon _{\mu \nu \alpha \beta }\epsilon
^{\nu }p^{\alpha }p^{\prime }{}^{\beta }.  \label{eq:MElem2}
\end{eqnarray}%
In Eq.\ (\ref{eq:MElem2}) $\ p$ and $\epsilon $ are the momentum and
polarization vector of the $T_{bc}^{0}$, $p^{\prime }$ is the momentum of
the scalar tetraquark $T_{cc}^{+}$. Here we also use the shorthand notations
$\overline{m}=m+m_{T}$ and $P_{\mu }=p_{\mu }^{\prime }+p_{\mu }$ with $%
m_{T} $ being the mass of the final-state tetraquark. The $q_{\mu }=p_{\mu
}-p_{\mu }^{\prime }$ is the momentum transferred to the leptons changing
within the limits $m_{l}^{2}\leq q^{2}\leq (m-m_{T})^{2}$, where $m_{l}$ is
the mass of the lepton $l$.

The form factors $G_{i}(q^{2})$ are key quantities to be extracted from the
sum rules. To this end, we consider the following three-point correlation
function:
\begin{eqnarray}
\Pi _{\mu \nu }(p,p^{\prime }) &=&i^{2}\int d^{4}xd^{4}ye^{i(p^{\prime
}y-px)}  \notag \\
&&\times \langle 0|\mathcal{T}\{J^{T}(y)J_{\mu }^{\mathrm{tr}}(0)J_{\nu
}^{^{\dagger }}(x)\}|0\rangle ,  \label{eq:CF7}
\end{eqnarray}%
where $J_{\nu }(x)$ and $J^{T}(y)$ are the interpolating currents
corresponding to the states $T_{bc}^{0}$ and $T_{cc}^{+}$, respectively. The
current $J_{\nu }(x)$ has been introduced by Eq.\ (\ref{eq:Curr}). The
interpolating current for the state$T_{cc}^{+}$ is given by the expression:
\begin{equation}
J^{T}(y)=\epsilon \widetilde{\epsilon }[c_{b}^{T}(y)C\gamma _{\alpha
}c_{c}(y)][\overline{u}_{d}(y)\gamma ^{\alpha }C\overline{d}_{e}^{T}(y)],
\label{eq:Curr2}
\end{equation}%
where $\epsilon \widetilde{\epsilon }=\epsilon ^{abc}\epsilon ^{ade}$. Here,
$\epsilon ^{abc}[c_{b}^{T}C\gamma _{\alpha }c_{c}]$ and $\epsilon ^{ade}[%
\overline{u}_{d}\gamma ^{\alpha }C\overline{d}_{e}^{T}]$ are the
axial-vector diquark and antidiquark, respectively. Then the scalar
designation of the final tetraquark $T_{cc}^{+}$ stems naturally from the
internal structure of the initial four-quark state $T_{bc}^{0}$, which is
the axial-vector particle composed of the scalar diquark $b^{T}C\gamma _{5}c$
and axial-vector antidiquark $\overline{u}\gamma _{\mu }C\overline{d}^{T}$.
The semileptonic decay $T_{bc}^{0}\rightarrow T_{cc}^{+}+W^{-}$ runs through
$b\rightarrow W^{-}c$, which transforms the scalar diquark $bc$ to the final
axial-vector $cc$, leaving, at the same time, unchanged the initial light
antidiquark; the light axial-vector antidiquark $\overline{u}\overline{d}$
appears both in the initial and final states. The designation of $T_{cc}^{+}$
as an axial-vector requires $\overline{u}\overline{d}$ to be a scalar, which
implies additional spin-rearrangement in the initial axial-vector $\overline{%
u}\overline{d}$ diquark, which evidently suppresses the corresponding
process.

Our strategy to derive sum rules for the form factors $G_{i}(q^{2})$ is the
same as in all of this kind studies. In fact, to determine the
phenomenological side of the sum rule $\Pi _{\mu \nu }^{\mathrm{Phys}%
}(p,p^{\prime })$ we express the correlation function $\Pi _{\mu \nu
}(p,p^{\prime })$ in terms of the spectroscopic parameters of particles
involving into the decay process. Afterwards we find the QCD side (or OPE)
side of the sum rules $\Pi _{\mu \nu }^{\mathrm{OPE}}(p,p^{\prime })$ by
computing the same correlation function in terms of quark propagators. By
matching the obtained results and utilizing the quark-hadron duality
assumption we extract sum rules and evaluate the physical quantities of
interest. Because the quark propagators contain quark, gluon and mixed
vacuum condensates, the sum rules express the physical quantities as
functions of nonperturbative parameters.

In the context of this approach the function $\Pi _{\mu \nu }^{\mathrm{Phys}%
}(p,p^{\prime })$ can be recast into the form%
\begin{eqnarray}
&&\Pi _{\mu \nu }^{\mathrm{Phys}}(p,p^{\prime })=\frac{\langle
0|J^{T}|T_{cc}^{+}(p^{\prime })\rangle \langle T_{cc}^{+}(p^{\prime
})|J_{\mu }^{\mathrm{tr}}|T_{bc}^{0}(p,\epsilon )\rangle }{%
(p^{2}-m^{2})(p^{\prime 2}-m_{T}^{2})}  \notag \\
&&\times \langle T_{bc}^{0}(p,\epsilon )|J_{\nu }^{^{\dagger }}|0\rangle
+\ldots ,  \label{eq:Phys1}
\end{eqnarray}%
where $m_{T}$ is the mass of $T_{cc}^{+}$. In the expression above we take
into account contribution appearing due to only the ground-state particles,
denoting contributions of the higher resonances and continuum states by the
dots.

Transformation of the ground-state term in $\Pi _{\mu \nu }^{\mathrm{Phys}%
}(p,p^{\prime })$ can be completed by detailing the matrix elements in its
expression. The matrix element of $T_{bc}^{0}$ and the matrix element for
the transition $\langle T_{cc}^{+}(p^{\prime })|J_{\mu }^{\mathrm{tr}%
}|T_{bc}^{0}(p,\epsilon )\rangle $ are given by Eqs.\ (\ref{eq:MElem1}) and (%
\ref{eq:MElem2}), respectively. The remaining quantity
\begin{equation}
\langle 0|J^{T}|T_{cc}^{+}(p^{\prime })\rangle =m_{T}f_{T},
\label{eq;MElem3}
\end{equation}%
has a simple form and depends only on the mass and coupling $f_{T}$ of the
tetraquark $T_{cc}^{+}$. Benefiting from these explicit formulas, for $\Pi
_{\mu \nu }^{\mathrm{Phys}}(p,p^{\prime },q^{2})$ we obtain
\begin{eqnarray}
&&\Pi _{\mu \nu }^{\mathrm{Phys}}(p,p^{\prime },q^{2})=\frac{fmf_{T}m_{T}}{%
(p^{2}-m^{2})(p^{\prime 2}-m_{T}^{2})}  \notag \\
&&\times \left\{ \overline{m}G_{0}(q^{2})\left( -g_{\mu \nu }+\frac{p_{\mu
}p_{\nu }}{m^{2}}\right) +\left[ \frac{G_{1}(q^{2})}{\overline{m}}P_{\mu
}\right. \right.  \notag \\
&&\left. +\frac{G_{2}(q^{2})}{\overline{m}}q_{\mu }\right] \left( -p_{\nu
}^{\prime }+\frac{m^{2}+m_{T}^{2}-q^{2}}{2m^{2}}p_{\nu }\right)  \notag \\
&&\left. -i\frac{G_{3}(q^{2})}{\overline{m}}\varepsilon _{\mu \nu \alpha
\beta }p^{\alpha }p^{\prime }{}^{\beta }\right\} +\ldots .  \label{eq:Phys2}
\end{eqnarray}

The function $\Pi _{\mu \nu }^{\mathrm{OPE}}(p,p^{\prime })$ forms the
second side of the sum rules:%
\begin{eqnarray}
&&\Pi _{\mu \nu }^{\mathrm{OPE}}(p,p^{\prime })=i^{2}\epsilon \widetilde{%
\epsilon }\int d^{4}xd^{4}ye^{i(p^{\prime }y-px)}\left( \mathrm{Tr}\left[
\gamma ^{\alpha }\widetilde{S}_{d}^{b^{\prime }e}(x-y)\right. \right.  \notag
\\
&&\left. \left. \times \gamma _{\nu }S_{u}^{a^{\prime }d}(x-y)\right] -%
\mathrm{Tr}\left[ \gamma ^{\alpha }\widetilde{S}_{d}^{a^{\prime
}e}(x-y)\gamma _{\nu }S_{u}^{b^{\prime }d}(x-y)\right] \right)  \notag \\
&&\times \left( \mathrm{Tr}\left[ \gamma _{\mu }(1-\gamma
_{5})S_{b}^{ia^{\prime }}(-x)\gamma _{5}\widetilde{S}_{c}^{bb^{\prime
}}(y-x)\gamma _{\alpha }S_{c}^{ci}(y)\right] \right.  \notag \\
&&\left. -\mathrm{Tr}\left[ \gamma _{\mu }(1-\gamma _{5})S_{b}^{ia^{\prime
}}(-x)\gamma _{5}\widetilde{S}_{c}^{cb^{\prime }}(y-x)\gamma _{\alpha
}S_{c}^{bi}(y)\right] \right) .  \notag \\
&&  \label{eq:QCD2}
\end{eqnarray}

The required sum rules for the form factors $G_{i}(q^{2})$ can be obtained
by equating invariant amplitudes corresponding to the same Lorentz
structures both in $\Pi _{\mu \nu }^{\mathrm{Phys}}(p,p^{\prime },q^{2})$
and $\Pi _{\mu \nu }^{\mathrm{OPE}}(p,p^{\prime })$. Because in the
three-point sum rules the invariant amplitudes are functions of $p^{\prime
2} $ and $p^{2}$, to suppress contributions of higher resonances and
continuum states we have to apply the double Borel transformation over these
variables. As a result, the final expressions depend on a set of Borel
parameters $\mathbf{M}^{2}=(M_{1}^{2},\ M_{2}^{2})$. The continuum
subtraction is performed in two channels using two continuum parameters $%
\mathbf{s}_{0}=(s_{0},\ s_{0}^{\prime })$. The form factor $G_{0}(q^{2})$ is
obtained by using the structure $g_{\mu \nu }$ and reads:
\begin{eqnarray}
&&G_{0}(\mathbf{M}^{2},\ \mathbf{s}_{0},~q^{2})=\frac{1}{\overline{m}%
fmf_{T}m_{T}}\int_{\mathcal{M}^{2}}^{s_{0}}ds\int_{4m_{c}^{2}}^{s_{0}^{%
\prime }}ds^{\prime }  \notag \\
&&\times \rho _{0}(s,s^{\prime
},q^{2})e^{(m^{2}-s)/M_{1}^{2}}e^{(m_{T}^{2}-s^{\prime })/M_{2}^{2}}.
\label{eq:FF0}
\end{eqnarray}%
The form factors $G_{i}(q^{2})$ ($i=1,2,3$) are derived employing other
Lorentz structures in the correlation functions:
\begin{eqnarray}
&&G_{i}(\mathbf{M}^{2},\ \mathbf{s}_{0},~q^{2})=\frac{\overline{m}}{%
fmf_{T}m_{T}}\int_{\mathcal{M}^{2}}^{s_{0}}ds\int_{4m_{c}^{2}}^{s_{0}^{%
\prime }}ds^{\prime }  \notag \\
&&\times \rho _{i}(s,s^{\prime
},q^{2})e^{(m^{2}-s)/M_{1}^{2}}e^{(m_{T}^{2}-s^{\prime })/M_{2}^{2}}.
\label{eq:FF}
\end{eqnarray}%
The sum rules (\ref{eq:FF0}) and (\ref{eq:FF}) are written down in terms of
the spectral densities $\rho _{i}(s,s^{\prime },q^{2})$ which are
proportional to the imaginary parts of the corresponding terms in $\Pi _{\mu
\nu }^{\mathrm{OPE}}(p,p^{\prime })$. They contain the perturbative and
nonperturbative contributions, and are calculated with dimension-5 accuracy.

\begin{table}[tbp]
\begin{tabular}{|c|c|}
\hline\hline
Quantity & Value \\ \hline\hline
$m_{T}$ & $(3845~\pm 175)~\mathrm{MeV}$ \\
$f_{T}$ & $(1.16\pm 0.26)\times 10^{-2}~\mathrm{GeV}^{4}$ \\
$m_{e}$ & $0.511~\mathrm{MeV}$ \\
$m_{\mu}$ & $105.658~\mathrm{MeV}$ \\
$m_{\tau}$ & $(1776.82~\pm 0.16)~\mathrm{MeV}$ \\
$G_F$ & $1.16637\times10^{-5}~\mathrm{GeV}^{-2}$ \\
$|V_{bc}|$ & $(42.2\pm 0.08)\times 10^{-3}$ \\ \hline\hline
\end{tabular}%
\caption{The mass and coupling of the final-state tetraquark $T_{cc}^{+}$
and other parameters used in numerical computations.}
\label{tab:Constants}
\end{table}

To compute the weak form factors $G_{i}(\mathbf{M}^{2},\ \mathbf{s}%
_{0},~q^{2})$ we need numerical values of parameters which enter to the sum
rules. The vacuum condensates are given in Eq.\ (\ref{eq:Parameters}),
whereas the spectroscopic parameters of the tetraquark $T_{cc}^{+}$ is
borrowed from our work \cite{Agaev:2019qqn}. The mass and coupling of the
initial particle $T_{bc}^{0}$ have been calculated in the previous section;
these and other parameters are collected in Table\ \ref{tab:Constants}. In
computations, we impose on the auxiliary parameters $\mathbf{M}^{2}$ and $%
\mathbf{s}_{0}$ the same constraints as in the mass calculations: the set ($%
M_{1}^{2},\ s_{0}$) for the initial particle channel is determined by Eq.\ (%
\ref{eq:Reg1}), whereas the set ($M_{2}^{2},\ s_{0}^{\prime }$) for $%
T_{cc}^{+}$ is chosen in the form \cite{Agaev:2019qqn}
\begin{equation}
M_{2}^{2}\in \lbrack 3,\ 4]~\mathrm{GeV}^{2},\ s_{0}^{\prime }\in \lbrack
19,\ 21]~\mathrm{GeV}^{2}.  \label{eq:Reg2}
\end{equation}

\begin{figure}[h]
\includegraphics[width=8.8cm]{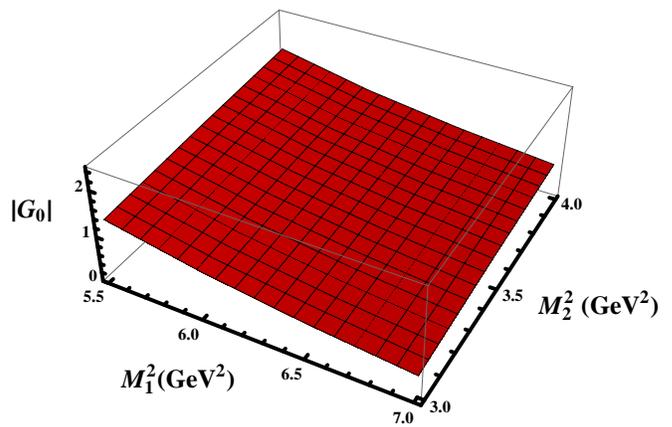}
\caption{ The form factor $|G_{0}|=|G_{0}(5~\mathrm{GeV}^{2})|$ as a
function of the Borel parameters $M_{1}^{2}$ and $M_{2}^{2}$ at $s_{0}=62~%
\mathrm{GeV}^{2}$ and $s_{0}^{\prime }=20~\mathrm{GeV}^{2}$.}
\label{fig:FFG0}
\end{figure}

Results of sum rule calculations in the case of $G_{0}(q^{2})$, as an
example, are shown in Fig.\ \ref{fig:FFG0}.
The similar predictions have been obtained for the remaining form factors as
well. The sum rule results for the functions $G_{i}(q^{2})$ are necessary,
but not enough to calculate the partial width of the process $%
T_{bc}^{0}\rightarrow T_{cc}^{+}l\overline{\nu }_{l}$. The reason is that
these form factors determine its differential decay rate $d\Gamma /dq^{2}$
(see, Appendix in Ref. \cite{Agaev:2018khe}). The partial width $\Gamma $
should be found by integrating $d\Gamma /dq^{2}$ over $q^{2}$ within limits
allowed by the kinematical constraints $m_{l}^{2}\leq q^{2}\leq
(m-m_{T})^{2} $. But sum rules do not cover all this region, and give
reliable results within the limits $m_{l}^{2}\leq q^{2}\leq 8\ ~\mathrm{GeV}%
^{2}$. Therefore, one has to introduce the model functions $F_{i}(q^{2})$,
which at accessible for the sum rule computations $q^{2}$ coincide with $%
G_{i}(q^{2})$, but can be extrapolated to the whole integration region.

The fit functions
\begin{equation}
F_{i}(q^{2})=F_{0}^{i}\exp \left[ c_{1}^{i}\frac{q^{2}}{m_{\mathrm{fit}}^{2}}%
+c_{2}^{i}\left( \frac{q^{2}}{m_{\mathrm{fit}}^{2}}\right) ^{2}\right] ,
\label{eq:FFunctions}
\end{equation}%
are convenient for these purposes. Here $F_{0}^{i},~c_{1}^{i},\ $ $c_{2}^{i}$
and $m_{\mathrm{fit}}^{2}$ are the fit parameters numerical values of which
are collected in Table \ref{tab:Fitparameters}.
\begin{table}[tbp]
\begin{tabular}{|c|c|c|c|c|}
\hline\hline
$F_i(q^2)$ & $F_{0}^{i}$ & $c_{1}^{i}$ & $c_{2}^{i}$ & $m^2_{\mathrm{fit}}\ (%
\mathrm{GeV}^2)$ \\ \hline\hline
$F_{0}(q^2)$ & $-0.92$ & $0.43$ & $-9.36$ & $50.48$ \\
$F_{1}(q^2)$ & $10.87$ & $2.83$ & $3.69$ & $50.48$ \\
$F_{2}(q^2)$ & $-2.61$ & $0.32$ & $4.44$ & $50.48$ \\
$F_{3}(q^2)$ & $-13.79$ & $2.06$ & $3.31$ & $50.48$ \\ \hline\hline
\end{tabular}%
\caption{The parameters of the fit functions $F_{i}(q^{2})$.}
\label{tab:Fitparameters}
\end{table}
Our predictions for the partial width of the semileptonic decay channels
are:
\begin{eqnarray}
\Gamma (T_{bc}^{0} &\rightarrow &T_{cc}^{+}e^{-}\overline{\nu }%
_{e})=(1.44\pm 0.35)\times 10^{-10}\ \mathrm{MeV},  \notag \\
\Gamma (T_{bc}^{0} &\rightarrow &T_{cc}^{+}\mu ^{-}\overline{\nu }_{\mu
})=(1.43\pm 0.34)\times 10^{-10}\ \mathrm{MeV},  \notag \\
\Gamma (T_{bc}^{0} &\rightarrow &T_{cc}^{+}\tau ^{-}\overline{\nu }_{\tau
})=(4.3\pm 1.1)\times 10^{-11}\ \mathrm{MeV}.  \notag \\
&&  \label{eq:Results}
\end{eqnarray}%
Results (\ref{eq:Results}) obtained in this section constitute an important
part of the full width of $T_{bc}^{0}$, and will be used below for its
evaluation.


\section{ Two-body weak decays $T_{bc}^{0}\rightarrow T_{cc}^{+}\protect\pi %
^{-}(K^{-},\ D^{-},\ D_{s}^{-})$}

\label{sec:Decays2}

The two-body weak decays $T_{bc}^{0}\rightarrow T_{cc}^{+}\pi ^{-}(K^{-},\
D^{-},\ D_{s}^{-})$ of the tetraquark $T_{bc}^{0}$ can be considered in the
context of the QCD factorization approach, which allows one to write
amplitudes and calculate widths of these processes. This method was
successfully applied to study two-body weak decays of the conventional
mesons \cite{Beneke:1999br,Beneke:2000ry}, and is used here to investigate
two-body decays of the tetraquark $T_{bc}^{0}$, when one of the final
particles is an exotic meson.

We consider in a detailed form only the decay $T_{bc}^{0}\rightarrow
T_{cc}^{+}\pi ^{-}$ , and write down final predictions for remaining
channels. At the quark level, the effective Hamiltonian for the this decay
is given by the expression
\begin{equation}
\widetilde{\mathcal{H}}^{\mathrm{eff}}=\frac{G_{F}}{\sqrt{2}}%
V_{bc}V_{ud}^{\ast }\left[ c_{1}(\mu )Q_{1}+c_{2}(\mu )Q_{2}\right] ,
\label{eq:EffHam}
\end{equation}%
where%
\begin{eqnarray}
Q_{1} &=&\left( \overline{d}_{i}u_{i}\right) _{\mathrm{V-A}}\left( \overline{%
c}_{j}b_{j}\right) _{\mathrm{V-A}},  \notag \\
Q_{2} &=&\left( \overline{d}_{i}u_{j}\right) _{\mathrm{V-A}}\left( \overline{%
c}_{j}b_{i}\right) _{\mathrm{V-A}},  \label{eq:Operators}
\end{eqnarray}%
and $i$ , $j$ are the color indices. Here $c_{1}(\mu )$ and $c_{2}(\mu )$
are the short-distance Wilson coefficients evaluated at the scale $\mu $ at
which the factorization is assumed to be correct. The shorthand notation $%
\left( \overline{q}_{1}q_{2}\right) _{\mathrm{V-A}}$ in Eq.\ (\ref%
{eq:Operators}) means
\begin{equation}
\left( \overline{q}_{1}q_{2}\right) _{\mathrm{V-A}}=\overline{q}_{1}\gamma
_{\mu }(1-\gamma _{5})q_{2}.  \label{eq:Not}
\end{equation}%
The amplitude of this decay can be written down in the following factorized
form%
\begin{eqnarray}
\mathcal{A} &=&\frac{G_{F}}{\sqrt{2}}V_{bc}V_{ud}^{\ast }a_{1}(\mu )\langle
\pi ^{-}(q)|\left( \overline{d}_{i}u_{i}\right) _{\mathrm{V-A}}|0\rangle
\notag \\
&&\times \langle T_{cc}^{+}(p^{\prime })|\left( \overline{c}_{j}b_{j}\right)
_{\mathrm{V-A}}|T_{bc}^{0}(p,\epsilon )\rangle  \label{eq:Amplitude}
\end{eqnarray}%
where
\begin{equation}
a_{1}(\mu )=c_{1}(\mu )+\frac{1}{N_{c}}c_{2}(\mu ),
\end{equation}%
and $N_{c}$ is the number of quark colors. The amplitude $\mathcal{A}$
describes the process in which the pion $\pi ^{-}$ is generated directly
from the color-singlet current $\left( \overline{d}_{i}u_{i}\right) _{%
\mathrm{V-A}}$. The matrix element $\langle T_{cc}^{+}(p^{\prime })|\left(
\overline{c}_{j}b_{j}\right) _{\mathrm{V-A}}|T_{bc}^{0}(p,\epsilon )\rangle $
has been introduced by Eq.\ (\ref{eq:MElem2}), whereas the matrix element of
the pion in given by the expression%
\begin{equation}
\langle \pi ^{-}(q)|\left( \overline{d}_{i}u_{i}\right) _{\mathrm{V-A}%
}|0\rangle =if_{\pi }q_{\mu },  \label{eq:ME4}
\end{equation}%
and is determined by its decay constant $f_{\pi }$.

Then, it is not difficult to see that $\mathcal{A}$ takes the form%
\begin{eqnarray}
&&\mathcal{A}=i\frac{G_{F}}{\sqrt{2}}f_{\pi }V_{bc}V_{ud}^{\ast }a_{1}(\mu
)(\epsilon p^{\prime })\left[ -\overline{m}G_{0}(q^{2})\right.  \notag \\
&&\left. +\frac{G_{1}(q^{2})}{\overline{m}}Pq+\frac{G_{2}(q^{2})}{\overline{m%
}}m_{\pi }^{2}\right] . 
 \label{eq:Amplitude2}
\end{eqnarray}%
The width of the decay $T_{bc}^{0}\rightarrow T_{cc}^{+}\pi ^{-}$ is
\begin{eqnarray}
&&\Gamma \left( T_{bc}^{0}\rightarrow T_{cc}^{+}\pi ^{-}\right) =\frac{%
G_{F}^{2}f_{\pi }^{2}}{48\pi m^{2}}|V_{bc}|^{2}|V_{ud}|^{2}a_{1}^{2}(\mu )
\notag \\
&&\times \lambda ^{3}(m^{2},m_{T}^{2},m_{\pi }^{2})\left[ \overline{m}%
^{2}|G_{0}|^{2}+\frac{|G_{1}|^{2}}{\overline{m}^{2}}(m^{2}-m_{T}^{2})^{2}%
\right.  \notag \\
&&+\frac{|G_{2}|^{2}}{\overline{m}^{2}}m_{\pi }^{4}-2\func{Re}\left[
G_{0}G_{1}^{\ast }\right] (m^{2}-m_{T}^{2})  \notag \\
&&\left. -2\func{Re}\left[ G_{0}G_{2}^{\ast }\right] m_{\pi }^{2}+2\func{Re}%
\left[ G_{1}G_{2}^{\ast }\right] \frac{(m^{2}-m_{T}^{2})m_{\pi }^{2}}{%
\overline{m}^{2}}\right] ,
\notag \\  \label{eq:NonLDW}
\end{eqnarray}%
where the weak form factors $G_{i}(q^{2})$ ($i=0,1,2$) are taken at $%
q^{2}=m_{\pi }^{2}$. In Eq. (\ref{eq:NonLDW}) the function $\lambda
(m^{2},m_{T}^{2},m_{\pi }^{2})$ is given by the formula
\begin{eqnarray}
&&\lambda \left( m^{2},m_{T}^{2},m_{\pi }^{2}\right) =\frac{1}{2m}\left[
m^{4}+m_{T}^{4}+m_{\pi }^{4}\right.  \notag \\
&&\left. -2\left( m^{2}m_{T}^{2}+m^{2}m_{\pi }^{2}+m_{T}^{2}m_{\pi
}^{2}\right) \right] ^{1/2}.  \label{eq:Lambda}
\end{eqnarray}%
The similar analysis can be performed for other decays $T_{bc}^{0}%
\rightarrow T_{cc}^{+}(K^{-},D^{-},D_{s}^{-})$ as well: relevant expressions
can by obtained from (\ref{eq:NonLDW}) using the spectroscopic parameters of
the mesons $K^{-},D^{-},$ and $D_{s}^{-}$, and by replacements $%
V_{ud}\rightarrow V_{us}$, $V_{cd}$, and $V_{cs}$, respectively.

Numerical computations can be carried out after fixing the spectroscopic
parameters of the final-state pseudoscalar mesons, weak form factors, and
CKM matrix elements. The masses and decay constants of the final-state
pseudoscalar mesons are presented in Table\ \ref{tab:MesonPar}. The weak
form factors $G_{i}(q^{2})$ ($i=0,1,2,3$), which are crucial parts of
calculations, have been obtained in the previous section. For CKM matrix
elements we use $|V_{ud}|=0.97420\pm 0.00021$, $|V_{us}|=0.2243\pm
0.0005 $, $|V_{cd}|=0.218\pm 0.004$ and $|V_{cs}|=0.997\pm 0.017$. The
values of the Wilson coefficients $c_{1}(m_{b}),\ $and $c_{2}(m_{b})$ with
next-to-leading order QCD corrections were presented in Refs.\ \cite%
{Buras:1992zv,Ciuchini:1993vr,Buchalla:1995vs}
\begin{equation}
c_{1}(m_{b})=1.117,\ c_{2}(m_{b})=-0.257.  \label{eq:WCoeff}
\end{equation}

\begin{table}[tbp]
\begin{tabular}{|c|c|}
\hline\hline
Quantity & Value \\ \hline\hline
$m_{\pi }$ & $139.570~\mathrm{MeV}$ \\
$m_{K}$ & $(493.677\pm 0.016)~\mathrm{MeV}$ \\
$m_{D}$ & $(1869.61 \pm 0.10)~\mathrm{MeV}$ \\
$m_{D_s}$ & $(1968.30\pm 0.11)~\textrm{MeV}$ \\
$f_{\pi }$ & $131~\text{MeV}$ \\
$f_{K}$ & $(155.72\pm 0.51)~\text{MeV}$ \\
$f_{D}$ & $(203.7 \pm 4.7)~\text{MeV}$ \\
$f_{D_s}$ & $(257.8 \pm 4.1)~\text{MeV}$ \\ \hline\hline
\end{tabular}%
\caption{Masses and decay constants of the pseudoscalar mesons.}
\label{tab:MesonPar}
\end{table}
For the decay $T_{bc}^{0}\rightarrow T_{cc}^{+}\pi ^{-}$, calculations lead
to the following result
\begin{eqnarray}
&&\Gamma \left( T_{bc}^{0}\rightarrow T_{cc}^{+}\pi ^{-}\right) =\left(
1.73\pm 0.38\right) \times 10^{-11}\ \mathrm{MeV}.  \notag \\
&&  \label{eq:NLDW1}
\end{eqnarray}%
Width of this decay is smaller than widths of the semileptonic decays, but
is comparable with them. For the remaining weak nonleptonic decays of the
tetraquark $T_{bc}^{0}$ we get
\begin{eqnarray}
&&\Gamma \left( T_{bc}^{0}\rightarrow T_{cc}^{+}K^{-}\right) =\left( 1.27\pm
0.26\right) \times 10^{-12}~\mathrm{MeV},  \notag \\
&&\Gamma \left( T_{bc}^{0}\rightarrow T_{cc}^{+}D^{-}\right) =\left( 1.65\pm
0.35\right) \times 10^{-12}~\mathrm{MeV},  \notag \\
&&\Gamma \left( T_{bc}^{0}\rightarrow T_{cc}^{+}D_{s}^{-}\right) =\left(
4.74\pm 0.99\right) \times 10^{-11}~\mathrm{MeV}.  \notag \\
&&  \label{eq:NLDW2}
\end{eqnarray}%
It is seen that partial widths only of the nonleptonic weak decays $%
T_{bc}^{0}\rightarrow T_{cc}^{+}D_{s}^{-}$ and $T_{bc}^{0}\rightarrow
T_{cc}^{+}\pi ^{-}$ are comparable with widths of the semileptonic modes \ (%
\ref{eq:Results}); contribution to the full width of $T_{bc}^{0}$ coming
from other two weak decays is neglidible.

Using Eqs.\ (\ref{eq:Results}) and (\ref{eq:NLDW2}), it is not difficult to
find the full width and mean lifetime of $T_{bc}^{0}$
\begin{eqnarray}
\Gamma _{\mathrm{full}} &=&(3.98\pm 0.51)\times 10^{-10}~\mathrm{MeV},
\notag \\
\tau  &=&1.65_{-0.18}^{+0.25}\times 10^{-12}~\mathrm{s}.  \label{eq:WL}
\end{eqnarray}%
Predictions for $\Gamma _{\mathrm{full}}$ and $\tau $ are among main results
of the present work.


\section{Strong decays $T_{bc}^{0}\rightarrow B^{\ast -}D^{+}$ and $%
T_{bc}^{0}\rightarrow \overline{B}^{\ast 0}D^{0}$}

\label{sec:Decays3}

Calculations of the mass of the tetraquark $T_{bc}^{0}$, performed in
Section \ref{sec:Mass}, due to uncertainties of the sum rule method do not
exclude also prediction $m=7260~\mathrm{MeV}$. In this scenario $T_{bc}^{0}$
is strong-interaction unstable particle and decays to conventional mesons $%
B^{\ast -}D^{+}$ and $\overline{B}^{\ast 0}D^{0}$. It is worth noting that $%
m=7260~\mathrm{MeV}$ is below the thresholds for strong decays $%
T_{bc}^{0}\rightarrow B^{-}D^{\ast +}$ and $T_{bc}^{0}\rightarrow \overline{B%
}^{0}D^{\ast 0}$, which forbids kinematically these processes. Below we
present in a detailed form our analysis of the decay $T_{bc}^{0}\rightarrow
B^{\ast -}D^{+}$ and provide final predictions for $T_{bc}^{0}\rightarrow
\overline{B}^{\ast 0}D^{0}$.

In the context of the QCD three-point sum rule method the strong decay $%
T_{bc}^{0}\rightarrow B^{\ast -}D^{+}$ can be studied using the correlation
function
\begin{eqnarray}
\widetilde{\Pi }_{\mu \nu }(p,p^{\prime }) &=&i^{2}\int
d^{4}xd^{4}ye^{i(p^{\prime }y-px)}\langle 0|\mathcal{T}\{J_{\mu }^{B^{\ast
}}(y)  \notag \\
&&\times J^{D}(0)J_{\nu }^{\dagger }(x)\}|0\rangle .  \label{eq:CF5}
\end{eqnarray}%
Here $J_{\nu }(x)$,$\ J^{D}(x)$ and $J_{\mu }^{B^{\ast }}(x)$ are the
interpolating currents for the tetraquark $T_{bc}^{0}$ and mesons $D^{+}$
and $B^{\ast -}$, respectively. The $J_{\nu }(x)$ is given by Eq.\ (\ref%
{eq:Curr}), whereas for the remaining two currents we use
\begin{equation}
J_{\mu }^{B^{\ast }}(x)=\overline{u}^{i}(x)\gamma _{\mu }b^{i}(x),\ \
J^{D}(x)=\overline{d}^{j}(x)i\gamma _{5}c^{j}(x).  \label{eq:Curr3}
\end{equation}%
The 4-momenta of the tetraquark $T_{bc}^{0}$ and meson $B^{\ast -}$ are $p$
and $p^{\prime }$, therefore, the momentum of the meson $D^{+}$ is $%
q=p-p^{\prime }$.

We follow the standard recipes and calculate the correlation function $%
\widetilde{\Pi }_{\mu \nu }(p,p^{\prime })$ using both the physical
parameters of the particles involved into the process, and quark
propagators. Separating the ground-state contribution from ones due to
higher resonances and continuum states,\ for the physical side of the sum
rule, we get%
\begin{eqnarray}
&&\widetilde{\Pi }_{\mu \nu }^{\mathrm{Phys}}(p,p^{\prime })=\frac{\langle
0|J_{\mu }^{B^{\ast }}|B^{\ast -}(p^{\prime },\epsilon ^{\prime })\rangle
\langle 0|J^{D}|D^{+}(q)\rangle }{(p^{\prime 2}-m_{B^{\ast
}}^{2})(q^{2}-m_{D}^{2})}  \notag \\
&&\times \frac{\langle D^{+}(q)B^{\ast -}(p^{\prime },\epsilon ^{\ast \prime
})|T_{bc}^{0}(p,\epsilon )\rangle \langle T_{bc}^{0}(p,\epsilon ^{\ast
})|J_{\nu }^{\dagger }|0\rangle }{(p^{2}-m^{2})}+\ldots  \notag \\
&&  \label{eq:CF6}
\end{eqnarray}

The function $\widetilde{\Pi }_{\mu \nu }^{\mathrm{Phys}}(p,p^{\prime })$
can be simplified by expressing the matrix elements in terms of the
tetraquark and mesons' physical parameters. The matrix element $\langle
T_{bc}^{0}(p,\epsilon ^{\ast })|J_{\nu }^{\dagger }|0\rangle $ can be found
using Eq.\ (\ref{eq:MElem1}). We introduce also the matrix elements of the
final-state mesons
\begin{eqnarray}
\langle 0|J^{D}|D^{+}\rangle &=&\frac{m_{D}^{2}f_{D}}{m_{c}},\   \notag \\
\langle 0|J_{\mu }^{B^{\ast }}|B^{\ast -}(p^{\prime },\epsilon ^{\prime
})\rangle &=&m_{B^{\ast }}f_{B^{\ast }}\epsilon _{\mu }^{\prime }.
\label{eq:Mel2}
\end{eqnarray}%
Here $m_{D}$,$\ m_{B^{\ast }}$ and $f_{D}$, $f_{B^{\ast }}$ are the masses
and decay constants of the mesons $D^{+}$ and $B^{\ast -}$, respectively. In
Eq.\ (\ref{eq:Mel2}) $\epsilon _{\mu }^{\prime }$ is the polarization vector
of the meson $B^{\ast -}$. We model $\langle D^{+}(q)B^{\ast -}(p^{\prime
},\epsilon ^{\ast \prime })|T_{bc}^{0}(p,\epsilon )\rangle $ in the form%
\begin{eqnarray}
&&\langle D^{+}(q)B^{\ast -}(p^{\prime },\epsilon ^{\ast \prime
})|T_{bc}^{0}(p,\epsilon )\rangle =g_{1}(q^{2})\left[ (p\cdot p^{\prime
})\right.  \notag \\
&&\left. \times (\epsilon \cdot \epsilon ^{\ast \prime })-(p\cdot \epsilon
^{\ast \prime })(p^{\prime }\cdot \epsilon )\right]  \label{eq:StVertex}
\end{eqnarray}%
and denote by $g_{1}(q^{2})$ the strong form factor corresponding to the
vertex $T_{bc}^{0}B^{\ast -}D^{+}.$ Then, it is not difficult to see that
\begin{eqnarray}
&&\widetilde{\Pi }_{\mu \nu }^{\mathrm{Phys}}(p,p^{\prime })=g_{1}\frac{%
m_{D}^{2}m_{B^{\ast }}mff_{D}f_{B^{\ast }}}{m_{c}(p^{2}-m^{2})(p^{\prime
2}-m_{B^{\ast }}^{2})}  \notag \\
&&\times \frac{1}{(q^{2}-m_{D}^{2})}\left[ \frac{1}{2}(m^{2}+m_{B^{\ast
}}^{2}-q^{2})g_{\mu \nu }-p_{\mu }p_{\nu }^{\prime }\right] +\ldots
\notag \\
\label{eq:Phys3}
\end{eqnarray}%
The correlation function $\widetilde{\Pi }_{\mu \nu }^{\mathrm{Phys}%
}(p,p^{\prime })$ has Lorentz structures proportional to $g_{\mu \nu }$ and $%
p_{\mu }p_{\nu }^{\prime }$. We work with the invariant amplitude $%
\widetilde{\Pi }^{\mathrm{Phys}}(p^{2},p^{\prime 2},q^{2})$ that corresponds
to the structure $\sim g_{\mu \nu }$. The double Borel transformation of
this amplitude over variables $p^{2}$ and $p^{\prime 2}$ forms the
phenomenological side of the sum rule.

To find the QCD side of the three-point sum rule, we calculate $\widetilde{%
\Pi }_{\mu \nu }(p,p^{\prime })$ in terms of the quark propagators and get
\begin{eqnarray}
&&\Pi _{\mu \nu }^{\mathrm{OPE}}(p,p^{\prime })=\int
d^{4}xd^{4}ye^{i(p^{\prime }y-px)}\left\{ \mathrm{Tr}\left[ \gamma
_{5}S_{c}^{jb}(-x)\gamma _{5}\right. \right.  \notag \\
&&\left. \times \widetilde{S}_{b}^{ia}(y-x)\gamma _{\mu }\widetilde{S}%
_{u}^{ai}(x-y)\gamma _{\nu }S_{d}^{bj}(x)\right] -\mathrm{Tr}\left[ \gamma
_{5}S_{c}^{jb}(-x)\right.  \notag \\
&&\left. \left. \times \gamma _{5}\widetilde{S}_{b}^{ia}(y-x)\gamma _{\mu }%
\widetilde{S}_{u}^{bi}(x-y)\gamma _{\nu }S_{d}^{aj}(x)\right] \right\} .
\label{eq:CF4}
\end{eqnarray}%
As in the case of the correlation function $\widetilde{\Pi }_{\mu \nu }^{%
\mathrm{Phys}}(p,p^{\prime })$ here, we also isolate the structure $\sim
g_{\mu \nu }$ and find the amplitude $\widetilde{\Pi }^{\mathrm{OPE}%
}(p^{2},p^{\prime 2},q^{2})$. The standard manipulations with invariant
amplitudes yield the following sum rule%
\begin{eqnarray}
&&g_{1}(q^{2})=\frac{2m_{c}}{m_{B^{\ast }}mff_{D}f_{B^{\ast }}}\frac{%
q^{2}-m_{D}^{2}}{m^{2}+m_{B^{\ast }}^{2}-q^{2}}  \notag \\
&&\times e^{m^{2}/M_{1}^{2}}e^{m_{B^{\ast }}^{2}/M_{2}^{2}}\widetilde{\Pi }^{%
\mathrm{OPE}}(\mathbf{M}^{2},\mathbf{s}_{0},q^{2}),  \label{eq:SRCoupl2}
\end{eqnarray}%
where $\mathbf{M}^{2}=(M_{1}^{2},M_{2}^{2}),$ and $\mathbf{s}%
_{0}=(s_{0},s_{0}^{\prime })$ are the Borel and continuum threshold
parameters. Apart from $q^{2}$, the form factor $g_{1}(q^{2})$ is also a
function of the Borel and continuum threshold parameters which, for
simplicity, are not shown explicitly in Eq.\ (\ref{eq:SRCoupl2}). The set $\
(M_{1}^{2},s_{0})$ corresponds to initial tetraquark channel, whereas $%
(M_{2}^{2},s_{0}^{\prime })$ describes the channel of the heavy final meson $%
B^{\ast -}$. Here, $\widetilde{\Pi }^{\mathrm{OPE}}(\mathbf{M}^{2},\mathbf{s}%
_{0},q^{2})$ is the invariant amplitude $\widetilde{\Pi }^{\mathrm{OPE}%
}(p^{2},p^{\prime 2},q^{2})$ after the double Borel transformation and
continuum subtraction procedures:%
\begin{eqnarray}
&&\widetilde{\Pi }^{\mathrm{OPE}}(\mathbf{M}^{2},\mathbf{s}%
_{0},q^{2})=\int_{(m_{b}+m_{c})^{2}}^{s_{0}}e^{-s/M_{1}^{2}}ds  \notag \\
&&\times \int_{m_{b}^{2}}^{s_{0}^{\prime }}ds^{\prime }e^{-s^{\prime
}/M_{2}^{2}}\rho \left( s,s^{\prime },q^{2}\right) .  \label{eq:InvAmp}
\end{eqnarray}%
The spectral density $\rho (s,s^{\prime },q^{2})$ is calculated as an
imaginary part of the relevant amplitude and contains the vacuum condensates
up to dimension 5.

The parameters, i.e., the vacuum condensates and masses of the $b$ and $c$
quarks, which are necessary for numerical computations are given by Eq. (\ref%
{eq:Parameters}). The mass and coupling of the tetraquark $T_{bc}^{0}$ have
been calculated in the present work. In computations we also use $%
m_{D^{0}}=(1864.84 \pm 0.07)~\mathrm{MeV}$ and $f_{D^{0}}=(203.7 \pm 4.7)~%
\mathrm{MeV}$, $m_{B^{\ast}}=(5325.2 \pm 0.4)~\mathrm{MeV}$ and $%
f_{B^{\ast}}=(210 \pm 6)~\mathrm{MeV}$, respectively. Parameters of the $D$
meson can be read out from Table \ref{tab:MesonPar}. The auxiliary
parameters for the $T_{bc}^{0}$ channel are chosen in accordance with Eq.\ (%
\ref{eq:Reg1}). For the set $(M_{2}^{2},s_{0}^{\prime })$ we use the regions%
\begin{equation}
M^{2}_{2}\in \lbrack 4.5,~5.5]~\mathrm{GeV}^{2},\ s_{0}^{\prime}\in \lbrack
32,~34]~\mathrm{GeV}^{2}.  \label{eq:Reg3}
\end{equation}

The sum rule method for $g_{1}(q^{2})$ gives reliable predictions only for $%
q^{2}<0$. Therefore, we introduce a variable $Q^{2}=-q^{2}$ and denote the
new function as $g_{1}(Q^{2})$. The width of the decay $T_{bc}^{0}%
\rightarrow B^{\ast -}D^{+}$ has to be computed using the strong form factor
at the mass shell of the $D^{+}$ meson $q^{2}=m_{D}^{2}$. This point is not
accessible to sum rule computations, but the problem can be solved by
employing a fit function $\mathcal{G}_{1}(Q^{2})$, which at the momenta $%
Q^{2}>0$ coincides with QCD sum rule predictions, but can be extrapolated to
the region of $Q^{2}<0$. Then, using the interpolating function $\mathcal{G}%
_{1}(Q^{2}),$one can find $g_{1}(-m_{D}^{2})$. The function $\mathcal{G}%
_{1}(Q^{2})$ does not differ from ones that we have used in Eq.\ (\ref%
{eq:FFunctions}), a difference being only in replacement of the fitting mass
with the mass of the tetraquark $m_{\mathrm{fit}}^{2}\rightarrow m^{2}$
\begin{equation}
\mathcal{G}_{1}(Q^{2})=\mathcal{G}_{0}^{1}\mathrm{\exp }\left[ \widetilde{c}%
_{1}^{1}\frac{Q^{2}}{m^{2}}+\widetilde{c}_{2}^{1}\left( \frac{Q^{2}}{m^{2}}%
\right) ^{2}\right] .  \label{eq:FitF}
\end{equation}%
The parameters $\mathcal{G}_{0}^{1}$, $\widetilde{c}_{1}^{1}$ and $%
\widetilde{c}_{2}^{1}$ have been fixed from numerical analyses $\mathcal{G}%
_{0}^{1}=1.11$, $\widetilde{c}_{1}^{1}=14.33,$ and $\widetilde{c}%
_{2}^{1}=-120.69$. This function at the mass shell $Q^{2}=-m_{D}^{2}$ gives%
\begin{equation}
g_{1}\equiv \mathcal{G}_{1}(-m_{D}^{2})=(0.25\pm 0.03)~\mathrm{GeV}^{-1}.
\label{eq:Coupl2}
\end{equation}%
The width of decay $T_{bc}^{0}\rightarrow B^{\ast -}D^{+}$ is determined by
the formula
\begin{equation}
\Gamma \lbrack T_{bc}^{0}\rightarrow B^{\ast -}D^{+}]=\frac{%
g_{1}^{2}m_{B^{\ast }}^{2}}{24\pi }\lambda \left( 3+2\frac{\lambda ^{2}}{%
m_{B^{\ast }}^{2}}\right) ,  \label{eq:DW2}
\end{equation}%
where $\lambda =\lambda \left( m^{2},m_{B^{\ast }}^{2},m_{D}^{2}\right) $.

Using Eqs.\ (\ref{eq:Coupl2}) and \ (\ref{eq:DW2}), one can easily calculate
the width of the decay$\ T_{bc}^{0}\rightarrow B^{\ast -}D^{+}$%
\begin{equation}
\Gamma \left[ T_{bc}^{0}\rightarrow B^{\ast -}D^{+}\right] =(31.1\pm 6.2)\
\mathrm{MeV}.  \label{eq:DW1Numeric}
\end{equation}

The second process $T_{bc}^{0}\rightarrow \overline{B}^{\ast 0}D^{0}$ can be
explored by the same manner. Here, we take into account that interpolating
currents have the following forms%
\begin{equation}
J_{\mu }^{\overline{B}^{\ast 0}}(x)=\overline{d}^{i}(x)\gamma _{\mu
}b^{i}(x),\ \ J^{D^{0}}(x)=\overline{u}^{j}(x)i\gamma _{5}c^{j}(x).
\label{eq:Curr4}
\end{equation}%
The remaining operations are standard manipulations in the context of the
sum rule method. Therefore, we do not see a necessity to provide a detailed
information on them. Let us note only that the fit function $\mathcal{G}%
_{2}(Q^{2})$ has the parameters $\mathcal{G}_{0}^{1}=1.11$, $\widetilde{c}%
_{1}^{2}=14.40$, and $\widetilde{c}_{2}^{2}=-121.11.$ At the mass shell of
the meson $D^{0}$ for the strong coupling we get
\begin{equation}
g_{2}\left( -m_{D^{0}}^{2}\right) =(0.26\pm 0.03)~\mathrm{GeV}^{-1},
\label{eq:Coupl3}
\end{equation}%
and%
\begin{equation}
\Gamma \lbrack T_{bc}^{0}\rightarrow \overline{B}^{\ast 0}D^{0}]=(32.4\pm
6.3)~\mathrm{MeV}.  \label{eq:DW2a}
\end{equation}%
Then, in the second scenario the full width of the axial-vector tetraquark $%
T_{bc}^{0}$ is%
\begin{equation}
\Gamma _{\mathrm{full}}=(63.5\pm 8.9)~\mathrm{MeV}.  \label{eq:W2}
\end{equation}%
This prediction for $\Gamma _{\mathrm{full}}$ is the main result obtained
utilizing the second option for $m$.


\section{Analysis and concluding notes}

\label{sec:Disc}
In the present work we have studied, in a rather detailed form, the
axial-vector tetraquark $T_{bc}^{0}$. As we have emphasized in Section \ref%
{sec:Int}, there are different predictions for its mass and stability
properties in the literature. We have calculated the mass $m$ and coupling $%
f $ of this tetraquark by means of the QCD sum rule method. Our result for $%
m $ does not allow us to solve unambiguously a problem with stability of the
tetraquark $T_{bc}^{0}$. Thus, the central value of the mass $7105~\mathrm{%
MeV}$ obtained in the present work is below both the strong and
electromagnetic thresholds, and therefore in this scenario $T_{bc}^{0}$ can
transform to conventional mesons only through the weak transitions. But
taking into account theoretical errors of computations and using the maximal
value of $m=7260~\mathrm{MeV}$, we see that $T_{bc}^{0}$ becomes unstable
against the strong and electromagnetic decays. We have explored both of
these scenarios and calculated the width and lifetime of $T_{bc}^{0}$.

In the framework of the first scenario, we have calculated the partial
widths of the semileptonic $T_{bc}^{0}\rightarrow T_{cc}^{+}l\overline{\nu }%
_{l}$ ($l=e,\mu $ and $\tau $) and two-body weak decays $T_{bc}^{0}%
\rightarrow T_{cc}^{+}\pi ^{-}(K^{-},\ D^{-},\ D_{s}^{-})$ of $T_{bc}^{0}$.
Using obtained information on these processes we have evaluated its full
width $\Gamma _{\mathrm{full}}=(3.98\pm 0.51)\times 10^{-10}\ \mathrm{MeV}$
and mean lifetime $\tau \approx 1.7~\mathrm{ps}$. In our previous work \cite%
{Sundu:2019feu} we computed the same parameters of the scalar tetraquark $%
Z_{bc}^{0}$. It is instructive to compare parameters of the scalar and
axial-vector $bc\overline{u}\overline{d}$ states with each other. The scalar
compound $Z_{bc}^{0}$ with the mass $6660~\mathrm{MeV}$ has a more stable
nature and lives $\tau \approx 21~\mathrm{ps}$ which is considerably longer
than $\tau \approx 1.7~\mathrm{ps}$ of the $T_{bc}^{0}$.

It is known that, the scalar tetraquark $T_{cc}^{+}$ decays strongly to a
pair of conventional $D^{+}D^{0}$ mesons \cite{Agaev:2019qqn}. Then, we can
estimate branching ratios of different weak decay channels of $T_{bc}^{0}$;
corresponding predictions are collected in Table\ \ref{tab:BRatio}.

\begin{table}[tbp]
\begin{tabular}{|c|c|}
\hline\hline
Channels & $\mathcal{BR}$ \\ \hline\hline
$D^{+}D^{0}e^{-}\overline{\nu }_{e}$ & $0.36$ \\
$D^{+}D^{0}\mu ^{-}\overline{\nu }_{\mu}$ & $0.36$ \\
$D^{+}D^{0}\tau ^{-}\overline{\nu }_{\tau }$ & $0.11$ \\
$D^{+}D^{0}\pi ^{-}$ & $0.043$ \\
$D^{+}D^{0}K^{-}$ & $0.003$ \\
$D^{+}D^{0} D^{-}$ & $0.004$ \\
$D^{+}D^{0}D_{s}^{-}$ & $0.12$ \\ \hline\hline
\end{tabular}%
\caption{The weak decay channels of the tetraquark $T_{bc}^{0}$ and
corresponding branching ratios.}
\label{tab:BRatio}
\end{table}

If mass of the tetraquark $T_{bc}^{0}$ is at around of $7260\ \mathrm{MeV}$,
it can decay strongly to conventional mesons. In present article we have
explored this scenario as well, and calculated partial widths of $S$-wave
decay channels $T_{bc}^{0}\rightarrow B^{\ast -}D^{+}$ and $%
T_{bc}^{0}\rightarrow \overline{B}^{\ast 0}D^{0}$. The full width $\Gamma _{%
\mathrm{full}}=(63.5\pm 8.9)\ \mathrm{MeV}$ of $T_{bc}^{0}$ estimated
employing these dominant decay modes characterizes $T_{bc}^{0}$ as a typical
unstable tetraquark. Branching ratios of the strong decay modes are equal to
\begin{eqnarray}
\mathcal{BR}(T_{bc}^{0} &\rightarrow &B^{\ast -}D^{+})\simeq 0.49,  \notag \\
\mathcal{BR}(T_{bc}^{0} &\rightarrow &\overline{B}^{\ast 0}D^{0})\simeq 0.51.
\label{eq:BR2}
\end{eqnarray}

Theoretical errors of sum rule computations and, as a result, different
predictions for the mass of the tetraquark $T_{bc}^{0}$ do not allow us to
interpret it unambiguously as strong- and electromagnetic-interaction stable
or unstable particle. The scenarios studied in our article provide useful
information on features of the axial-vector tetraquark $T_{bc}^{0}$ and may
be useful for its experimental and theoretical investigations.


\begin{thebibliography}{99}

\bibitem{Ader:1981db} J.~P.~Ader, J.~M.~Richard, and P.~Taxil,
Phys.\ Rev.\ D \textbf{25}, 2370 (1982).


\bibitem{Lipkin:1986dw} H.~J.~Lipkin,
Phys.\ Lett.\ B \textbf{172}, 242 (1986).


\bibitem{Zouzou:1986qh} S.~Zouzou, B.~Silvestre-Brac, C.~Gignoux, and
J.~M.~Richard, 
Z.\ Phys.\ C \textbf{30}, 457 (1986).


\bibitem{Carlson:1987hh} J.~Carlson, L.~Heller, and J.~A.~Tjon,
Phys.\ Rev.\ D \textbf{37}, 744 (1988).


\bibitem{Janc:2004qn} D.~Janc and M.~Rosina,
Few Body Syst.\ \textbf{35}, 175 (2004).


\bibitem{Cui:2006mp} Y.~Cui, X.~L.~Chen, W.~Z.~Deng, and S.~L.~Zhu,
HEPNP \textbf{31}, 7 (2007).


\bibitem{Vijande:2006jf} J.~Vijande, A.~Valcarce, and K.~Tsushima,
Phys.\ Rev.\ D \textbf{74}, 054018 (2006).


\bibitem{Ebert:2007rn} D.~Ebert, R.~N.~Faustov, V.~O.~Galkin, and W.~Lucha,
Phys.\ Rev.\ D \textbf{76}, 114015 (2007).


\bibitem{Navarra:2007yw} F.~S.~Navarra, M.~Nielsen, and S.~H.~Lee,
Phys.\ Lett.\ B \textbf{649}, 166 (2007).


\bibitem{Du:2012wp} M.~L.~Du, W.~Chen, X.~L.~Chen, and S.~L.~Zhu,
Phys.\ Rev.\ D \textbf{87}, 014003 (2013).


\bibitem{Hyodo:2012pm} T.~Hyodo, Y.~R.~Liu, M.~Oka, K.~Sudoh, and S.~Yasui,
Phys.\ Lett.\ B \textbf{721}, 56 (2013).


\bibitem{Esposito:2013fma} A.~Esposito, M.~Papinutto, A.~Pilloni,
A.~D.~Polosa, and N.~Tantalo,
Phys.\ Rev.\ D \textbf{88}, 054029 (2013).


\bibitem{Aaij:2017ueg} R.~Aaij \textit{et al.} [LHCb Collaboration],
Phys.\ Rev.\ Lett.\ \textbf{119}, 112001 (2017).


\bibitem{Karliner:2017qjm} M.~Karliner and J.~L.~Rosner,
Phys.\ Rev.\ Lett.\ \textbf{119}, 202001 (2017).


\bibitem{Eichten:2017ffp} E.~J.~Eichten and C.~Quigg,
Phys.\ Rev.\ Lett.\ \textbf{119}, 202002 (2017).


\bibitem{Agaev:2018khe} S.~S.~Agaev, K.~Azizi, B.~Barsbay, and H.~Sundu,
Phys.\ Rev.\ D \textbf{99}, 033002 (2019).


\bibitem{Feng:2013kea} G.-Q.~Feng, X.-H.~Guo, and B.-S.~Zou,
arXiv:1309.7813 [hep-ph]. 


\bibitem{Francis:2018jyb} A.~Francis, R.~J.~Hudspith, R.~Lewis, and
K.~Maltman,
arXiv:1810.10550 [hep-lat]. 


\bibitem{Caramees:2018oue} T.~F.~Carames, J.~Vijande, and A.~Valcarce,
arXiv:1812.08991 [hep-ph]. 


\bibitem{Sundu:2019feu} H.~Sundu, S.~S.~Agaev, and K.~Azizi,
arXiv:1903.05931 [hep-ph].


\bibitem{Sundu:2018uyi} H.~Sundu, B.~Barsbay, S.~S.~Agaev, and K.~Azizi,
Eur.\ Phys.\ J.\ A \textbf{54}, 124 (2018).


\bibitem{Agaev:2019qqn} S.~S.~Agaev, K.~Azizi, and H.~Sundu,
arXiv:1903.11975 [hep-ph].


\bibitem{Beneke:1999br} M.~Beneke, G.~Buchalla, M.~Neubert, and
C.~T.~Sachrajda,
Phys.\ Rev.\ Lett.\ \textbf{83}, 1914 (1999).


\bibitem{Beneke:2000ry} M.~Beneke, G.~Buchalla, M.~Neubert, and
C.~T.~Sachrajda,
Nucl.\ Phys.\ B \textbf{591}, 313 (2000).


\bibitem{Buras:1992zv} A.~J.~Buras, M.~Jamin, and M.~E.~Lautenbacher, Nucl.\
Phys.\ B \textbf{400}, 75 (1993).


\bibitem{Ciuchini:1993vr} M.~Ciuchini, E.~Franco, G.~Martinelli, and
L.~Reina, Nucl.\ Phys.\ B \textbf{415}, 403 (1994).


\bibitem{Buchalla:1995vs} G.~Buchalla, A.~J.~Buras, and M.~E.~Lautenbacher,
Rev.\ Mod.\ Phys.\ \textbf{68}, 1125 (1996).
\end{thebibliography}
\end{document}